# Quantum Simulation of Antiferromagnetic Heisenberg Chain with Gate-Defined Quantum Dots

C. J. van Diepen,[1,†] T.-K. Hsiao,[1,†] U. Mukhopadhyay,[1] C. Reichl,[2] W. Wegscheider,[2] and L. M. K. Vandersypen[1,*]

[1]*QuTech and Kavli Institute of Nanoscience, Delft University of Technology, 2600 GA Delft, Netherlands*
[2]*Solid State Physics Laboratory, ETH Zürich, 8093 Zürich, Switzerland*

Quantum-mechanical correlations of interacting fermions result in the emergence of exotic phases. Magnetic phases naturally arise in the Mott-insulator regime of the Fermi-Hubbard model, where charges are localized and the spin degree of freedom remains. In this regime, the occurrence of phenomena such as resonating valence bonds, frustrated magnetism, and spin liquids is predicted. Quantum systems with engineered Hamiltonians can be used as simulators of such spin physics to provide insights beyond the capabilities of analytical methods and classical computers. To be useful, methods for the preparation of intricate many-body spin states and access to relevant observables are required. Here, we show the quantum simulation of magnetism in the Mott-insulator regime with a linear quantum-dot array. We characterize the energy spectrum for a Heisenberg spin chain, from which we can identify when the conditions for homogeneous exchange couplings are met. Next, we study the multispin coherence with global exchange oscillations in both the singlet and triplet subspace of the Heisenberg Hamiltonian. Last, we adiabatically prepare the low-energy global singlet of the homogeneous spin chain and probe it with two-spin singlet-triplet measurements on each nearest-neighbor pair and the correlations therein. The methods and control presented here open new opportunities for the simulation of quantum magnetism benefiting from the flexibility in tuning and layout of gate-defined quantum-dot arrays.

## I. INTRODUCTION

Analog quantum simulations of magnetism [1] have been performed with a rich variety of experimental platforms, ranging from ultracold atoms in optical lattices [2–7] to trapped ions [8,9], scanning tunneling microscopy of atoms on metallic surfaces [10], and superconducting circuits [11]. A recent addition is the use of gate-defined quantum dots as a platform for quantum simulation of the Fermi-Hubbard model. The abilities to independently control the filling of the array, the local electrochemical potentials, and the hopping energy between sites are complemented with methods to probe the charge configuration across an array, spin states, and the electrical susceptibility as well as transport through the system [12]. These already enabled the observation of the transition from Coulomb blockade to collective Coulomb blockade [13] and the observation of Nagaoka ferromagnetism [14], a form of purely itinerant ferromagnetism which occurs at doping with a single hole.

In the Mott-insulator regime, where all sites are occupied by one electron, magnetism is governed by the Heisenberg exchange interaction [15], which favors antiferromagnetic spin alignment. Earlier studies on quantum-dot arrays in this regime demonstrate the sequential control of exchange couplings enabling coherent state transfer [16] and a method to handle cross talk in simultaneous control of exchange couplings [17]. In order to study the many-body properties of this system, novel methods are needed for state preparation in the presence of disorder and temperature as well as for probing spin correlations. Preparation of the Heisenberg ground state is both a useful and exciting goal because of its potential applications such as quantum information transfer [18–22] and quantum simulation of magnetic phases [23–25].

In this work, we simulate the antiferromagnetic Heisenberg spin chain in a gate-defined quadruple quantum dot. For this purpose, we develop experimental techniques based on energy spectroscopy and coherent oscillations of the global spin state. These include methods for many-body spin-state preparation and singlet-triplet correlation measurements, which form a powerful probe for the characterization of a many-body spin state [26]. We use these methods to engineer a chain with homogeneous exchange couplings. Finally, we adiabatically prepare the low-energy singlet eigenstate of the homogeneous Heisenberg chain and characterize the state with single-shot singlet-triplet readout on all nearest-neighbor pairs.

*Corresponding author.
l.m.k.vandersypen@tudelft.nl
†These authors contributed equally to this work.

TABLE I.  Four-spin shared eigenstates of $\hat{S}^2$ and $\hat{S}^z$ expressed in a basis of two-spin singlets and triplets on either the left and right or middle and outer pair. States in the rightmost column are the same as states on the same row in the column to the left of it. The four-spin states shown here are, in general, not eigenstates of the Heisenberg Hamiltonian, but the Hamiltonian does operate within a specific $(S, m_S)$ subspace.

| State | $(S, m_S)$ | Left- and right-pair basis | Middle- and outer-pair basis |
|---|---|---|---|
| $Q^{\pm\pm}$ | $(2,\pm 2)$ | $|T_{12}^{\pm}T_{34}^{\pm}\rangle$ | $|T_{23}^{\pm}T_{14}^{\pm}\rangle$ |
| $Q^{\pm}$ | $(2,\pm 1)$ | $(1/\sqrt{2})(|T_{12}^0 T_{34}^{\pm}\rangle + |T_{12}^{\pm}T_{34}^0\rangle)$ | $(1/\sqrt{2})(|T_{23}^0 T_{14}^{\pm}\rangle + |T_{23}^{\pm}T_{14}^0\rangle)$ |
| $Q^0$ | $(2,0)$ | $(1/\sqrt{6})(|T_{12}^+ T_{34}^-\rangle + |T_{12}^- T_{34}^+\rangle + 2|T_{12}^0 T_{34}^0\rangle)$ | $(1/\sqrt{6})(|T_{23}^+ T_{14}^-\rangle + |T_{23}^- T_{14}^+\rangle + 2|T_{23}^0 T_{14}^0\rangle)$ |
| $T_k^{\pm}$ | $(1,\pm 1)$ | $|2_{T^{\pm}}\rangle = (1/\sqrt{2})(|T_{12}^0 T_{34}^{\pm}\rangle - |T_{12}^{\pm}T_{34}^0\rangle)$ | $(1/\sqrt{2})(|S_{23}T_{14}^{\pm}\rangle + |T_{23}^{\pm}S_{14}\rangle)$ |
|  |  | $|1_{T^{\pm}}\rangle = |T_{12}^{\pm}S_{34}\rangle$ | $\frac{1}{2}(|S_{23}T_{14}^{\pm}\rangle - |T_{23}^{\pm}S_{14}\rangle - |T_{23}^0 T_{14}^{\pm}\rangle - |T_{23}^{\pm}T_{14}^0\rangle)$ |
|  |  | $|0_{T^{\pm}}\rangle = |S_{12}T_{34}^{\pm}\rangle$ | $\frac{1}{2}(|S_{23}T_{14}^{\pm}\rangle - |T_{23}^{\pm}S_{14}\rangle + |T_{23}^0 T_{14}^{\pm}\rangle - |T_{23}^{\pm}T_{14}^0\rangle)$ |
| $T_k^0$ | $(1,0)$ | $(1/\sqrt{2})(|T_{12}^+ T_{34}^-\rangle - |T_{12}^- T_{34}^+\rangle)$ | $(1/\sqrt{2})(|S_{23}T_{14}^0\rangle + |T_{23}^0 S_{14}\rangle)$ |
|  |  | $|T_{12}^0 S_{34}\rangle$ | $\frac{1}{2}(|S_{23}T_{14}^0\rangle - |T_{23}^0 S_{14}\rangle - |T_{23}^+ T_{14}^-\rangle + |T_{23}^- T_{14}^+\rangle)$ |
|  |  | $|S_{12}T_{34}^0\rangle$ | $\frac{1}{2}(|S_{23}T_{14}^0\rangle - |T_{23}^0 S_{14}\rangle + |T_{23}^+ T_{14}^-\rangle - |T_{23}^- T_{14}^+\rangle)$ |
| $S_k$ | $(0,0)$ | $|1_S\rangle = (1/\sqrt{3})(|T_{12}^+ T_{34}^-\rangle + |T_{12}^- T_{34}^+\rangle - |T_{12}^0 T_{34}^0\rangle)$ | $(\sqrt{3}/2)|S_{23}S_{14}\rangle + (1/2\sqrt{3})(|T_{23}^0 T_{14}^0\rangle - |T_{23}^+ T_{14}^-\rangle - |T_{23}^- T_{14}^+\rangle)$ |
|  |  | $|0_S\rangle = |S_{12}S_{34}\rangle$ | $\frac{1}{2}(|S_{23}S_{14}\rangle - |T_{23}^0 T_{14}^0\rangle + |T_{23}^+ T_{14}^-\rangle + |T_{23}^- T_{14}^+\rangle)$ |

## II. HEISENBERG SPIN CHAIN

The Heisenberg isotropic exchange Hamiltonian, while giving rise to rich emergent phenomena, has a simple form:

$$H_{\text{Heis}} = \sum_{\langle i,j\rangle} J_{ij}\left(\vec{S}_i \cdot \vec{S}_j - \frac{1}{4}\right), \qquad (1)$$

with $J_{ij}$ the exchange coupling between spins on sites $i$ and $j$, $\vec{S}_i$ the vector of spin operators for site $i$, and the summation over nearest neighbors only. The conventional $-\frac{1}{4}$ offset ascertains that the two-spin triplets have zero energy contribution in the absence of an external field. For quantum-dot systems, the exchange coupling is typically positive [27]; thus, neighboring spins prefer to antialign or, more precisely, tend to form local singlets. In addition, a Zeeman splitting can be induced with an external magnetic field, which energetically splits spin states according to their magnetization as

$$H_{\text{ext}} = g\mu_B B_{\text{ext}}\sum_i \hat{S}_i^z, \qquad (2)$$

with $g$ the Landé $g$ factor, $\mu_B$ the Bohr magneton, and $B_{\text{ext}}$ the external magnetic field.

The properties of a Heisenberg spin chain have theoretically been studied extensively, with as most famous result the exact solution of the energy spectrum and eigenstates of the homogeneous chain using the Bethe ansatz [28]. Intuitive insights can be obtained from the symmetries of the Heisenberg Hamiltonian, due to which the Hilbert space can be separated into subspaces, which are eigenspaces for the total spin operator $\hat{S}^2$, with eigenvalues $S(S+1)$, and the spin operator in the $z$ direction, $\hat{S}^z$, with eigenvalues $m_S \in [-S, \ldots, S]$. The dimensions for these subspaces can be obtained from the Clebsch-Gordan decomposition. For four spins, this results in two global singlet states for which $S = 0$, nine global triplet states for which $S = 1$, and five states with $S = 2$, which form a quintuplet. The triplets are separated into three three-dimensional subspaces and are denoted by $T_k^\alpha$, with the subspace magnetization $\alpha \in [-, 0, +]$ and $k$ labeling the energy level, where $k = 0$ for the state with lowest energy in the respective subspace. Similarly, the singlets are denoted by $S_k$, and the quintuplets by $Q^\beta$ with $\beta \in [--, -, 0, +, ++]$.

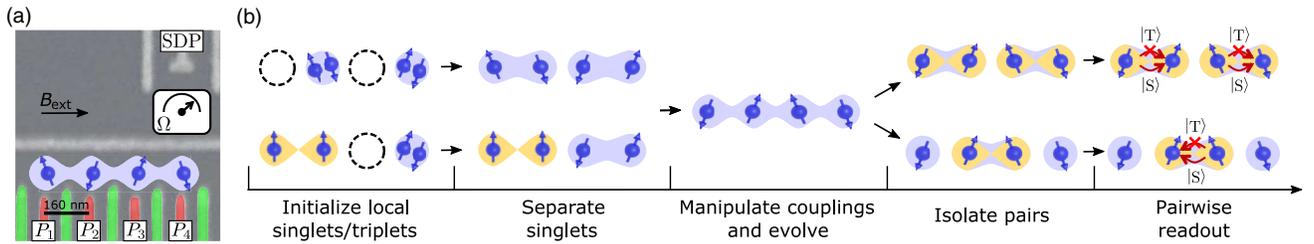

FIG. 1.  Device and spin-chain operation. (a) False-colored scanning electron micrograph of a device nominally identical to the one used for the experiments. The resistance meter, indicated with $\Omega$, shows the location of the sensing dot. Plunger gates for the dots are colored in red and labeled with $P_i$, and the plunger for the sensing dot is labeled with SDP. (b) Schematic illustration of the experimental sequence for the spin chain, consisting of five stages: initialization of local singlets or triplets, separation of singlets, manipulation of exchange couplings and time evolution, isolation of either the left and the right pair or the middle pair by switching off specific exchange couplings, and pairwise singlet-triplet readout.

The global spin states can be characterized in terms of the probabilities to measure either two-spin singlets $|S_{ij}\rangle = (1/\sqrt{2})(|\uparrow_i \downarrow_j\rangle - |\downarrow_i \uparrow_j\rangle)$ or triplets $|T_{ij}^+\rangle = |\uparrow_i \uparrow_j\rangle$, $|T_{ij}^0\rangle = (1/\sqrt{2})(|\uparrow_i \downarrow_j\rangle + |\downarrow_i \uparrow_j\rangle)$, and $|T_{ij}^-\rangle = |\downarrow_i \downarrow_j\rangle$, where $i$ and $j$ indicate the site. The simultaneous eigenstates of $\hat{S}^2$ and $\hat{S}^z$ for four-spin states can be expressed in this pairwise singlet-triplet basis as shown in Table I. Appendix A discusses the limitations of singlet-triplet measurements to distinguish spin states.

Alternatively, we can characterize the Heisenberg spin chain via its energy spectrum. Based on the symmetries of the Hamiltonian, for four spins the global singlet states form a two-dimensional subspace. For this subspace, the Heisenberg Hamiltonian is

$$H_{(0,0)} = \begin{pmatrix} -J_{12} - \frac{1}{4}J_{23} - J_{34} & \frac{\sqrt{3}}{4}J_{23} \\ \frac{\sqrt{3}}{4}J_{23} & -\frac{3}{4}J_{23} \end{pmatrix}, \qquad (3)$$

with basis states $|0_S\rangle$ and $|1_S\rangle$ from Table I. This subspace has been proposed as a singlet-only exchange-only qubit implementation, which offers increased coherence due to the reduced influence of nuclear spins [29].

For the global triplet states, the three three-dimensional subspaces are identical in terms of energy splittings. The Heisenberg Hamiltonian for each of these triplet subspaces is

$$H_{(1,1)} = \begin{pmatrix} -J_{12} - \frac{1}{4}J_{23} & -\frac{1}{4}J_{23} & -\frac{1}{2\sqrt{2}}J_{23} \\ -\frac{1}{4}J_{23} & -\frac{1}{4}J_{23} - J_{34} & -\frac{1}{2\sqrt{2}}J_{23} \\ -\frac{1}{2\sqrt{2}}J_{23} & -\frac{1}{2\sqrt{2}}J_{23} & -\frac{1}{2}J_{23} \end{pmatrix}, \quad (4)$$

with basis $|0_{T^+}\rangle$, $|1_{T^+}\rangle$, and $|2_{T^+}\rangle$ from Table I. The quintuplet states have zero energy contribution from the Heisenberg Hamiltonian but can be energetically split with an external magnetic field.

The energy differences in the subspaces reveal information about the exchange coupling strengths, and characteristic features can be identified. For the singlet subspace, the energy splitting is $\frac{1}{2}\sqrt{(2J_{12} + 2J_{34} - J_{23})^2 + 3J_{23}^2}$. It follows that, given $J_{12} = J_{34}$, the energy splitting is minimized when $J_{23} = J_{12} = J_{34}$ and, thus, for homogeneous exchange couplings. For the triplet subspace, the energy difference between the two lowest-energy states is minimized if $J_{12} = J_{34}$. If $J_{12} = J_{23} = J_{34}$, the triplet states are equally spaced in energy (see Appendix B for simulated energy diagrams). These characteristic features for the energy spectrum of the Heisenberg Hamiltonian will be experimentally identified, but first we introduce the quantum-dot device and the experimental operation.

## III. DEVICE AND EXPERIMENTAL OPERATION

The prototype for the simulation of an antiferromagnetic Heisenberg spin chain consists of a quadruple dot and a sensing dot, which are formed in a device nominally identical to that shown in Fig. 1(a) (see Appendix C 1 for details). The device is based on a GaAs/AlGaAs heterostructure, since this is the technology in which we are able to fabricate high-quality and well-controlled quantum-dot arrays. The exchange couplings are induced by electron wave function overlap, which we here control by detuning the potentials of neighboring dots, such that one electron shifts toward the other [30] (we note that independent control of the exchange couplings can also be achieved by adjusting the tunnel couplings [13,17,31]). In order to control the detuning between one pair of dots without affecting the detuning between other pairs, we define the detunings $\varepsilon_{ij}$ as

$$\begin{pmatrix} \varepsilon_{12} \\ \varepsilon_{23} \\ \varepsilon_{34} \end{pmatrix} = \begin{pmatrix} -1 & 1 & 1 & 1 \\ 1 & 1 & -1 & -1 \\ -1 & -1 & -1 & 1 \end{pmatrix} \begin{pmatrix} \varepsilon_1 \\ \varepsilon_2 \\ \varepsilon_3 \\ \varepsilon_4 \end{pmatrix}, \qquad (5)$$

where $\varepsilon_i$ is the negative local energy offset for site $i$ and $\varepsilon_{ij} = 0$ at the interdot transition between charge occupations (1111) and (0211), (1201), and (1102) for $\varepsilon_{12}$, $\varepsilon_{23}$, and $\varepsilon_{34}$, respectively. The $\varepsilon_i$ are independently controlled using virtual plunger gates, which are linear combinations of the voltages applied to the gates $P_i$ [13] (in the figures below, we express $\varepsilon_{ij}$ in units of mV; in Appendix C 2, we specify the conversion factor between energy and the applied voltage). The dependence of exchange couplings on detunings can be modeled as [32]

$$J_{ij} = \frac{1}{2}\left(\varepsilon_{ij} + \sqrt{8t_{ij}^2 + \varepsilon_{ij}^2}\right), \qquad (6)$$

with $t_{ij}$ the tunnel coupling between dots $i$ and $j$. In this way, the exchange couplings can be set independently with the detunings, and increasing detuning results in increasing exchange strength. This method can be extended to larger chains based on alternating the detuning directions such as defined in Eq. (5), which prevents unwanted charge transitions. In addition, the amount of change in chemical potentials could, in the future, be further reduced, by leveraging the fact that the effect of detuning on exchange is weaker closer to the charge symmetry point.

The experimental sequence used to operate the quantum-dot spin chain is schematically depicted in Fig. 1(b) and is described step by step here (charge-stability diagrams and sequence details are provided in Appendix C 1). Initially, the quadruple dot is tuned in either the (0202) or (1102) charge occupation. For the (0202) case, we load local singlets in the second and fourth dots, by allowing tunneling or cotunneling between dots and the reservoirs. For the (1102) configuration, we load a thermal mixture of two-spin states on the left pair, postselect for triplet loading (see Appendix C 4), and load a singlet on dot four. Next, the electrons are separated to obtain (1111) charge occupation. The global spin state remains a product of local spin pairs, because the left- and right-pair exchange remain large compared to the hyperfine field from the nuclear spins, and the middle-pair exchange coupling is kept small. Then, during the manipulation stage, the exchange couplings are diabatically or adiabatically changed by applying gate voltage pulses with variable rise time. In this work, the pulses are always diabatic with respect to anticrossings between states with different magnetization. Subsequently, the four-spin state evolves under the newly set exchange couplings. Finally, the spin pair(s) to be measured is(are) diabatically isolated from the other spins and measured with single-shot singlet-triplet readout based on Pauli spin blockade [33]. Here, either the left and right pair are sequentially read out, while parking the other pair to avoid capacitive cross talk [34], or the middle pair is read out.

In the remainder of this work, we focus on realizing homogeneous exchange couplings throughout the spin chain. In principle, this could be achieved by calibrating the exchange couplings one at a time [17] and extrapolating to the required tuning while compensating for cross talk. Instead, we develop a two-step spectroscopy method from which we identify directly when the condition of homogeneous exchange couplings is met.

## IV. ENERGY SPECTROSCOPY

For gate-defined quantum dots, information about the energy-level spectrum can be obtained from the degeneracies between spin states with different magnetization. This so-called spin funnel method has been used extensively in quantum-dot arrays of various lengths [30,35,36]. Here, we use the same underlying principles in a novel method for simultaneously characterizing multiple exchange coupling strengths in the spin chain. In addition, since the system size is small enough to allow classical numerical computation of its energy-level spectrum, we can validate the quantum simulator by comparing the measured energy spectrum to the numerically computed spectrum.

For the energy spectroscopy measurements, we prepare spin singlets on the left and right dot pairs [see Fig. 1(b)] (i.e., we prepare in the low-energy global singlet $|0_S\rangle$), diabatically pulse the exchange couplings, allow the system to evolve for 100 ns, and read out the left and right pair. The duration of 100 ns is chosen to allow the coherent time evolution kick started by the pulse to largely damp out. This measurement gives access to correlations in the singlet-triplet occupations, $P_{ST}$, $P_{TT}$, $P_{TS}$, and $P_{SS}$, where the left (right) subindex corresponds to the left- (right-) pair outcome. Decreased $P_{SS}$ indicates mixing of the low-energy global singlet with one of the triplet or quintuplet states. Such mixing occurs most manifestly at anticrossings between the low-energy global singlet state and the polarized states with $m_S = 1, 2$, induced by the gradients of the hyperfine field and the spin-orbit interaction. Depending on which of the other probabilities increases, we can infer information on the nature of the polarized state involved in that specific anticrossing.

We now examine and interpret the spectra in detail. For the measurement shown in Fig. 2(a), the left- and right-pair detunings during the manipulation stage are varied in the presence of a 40 mT magnetic field. The middle-pair detuning is kept fixed and such that the middle exchange coupling is small compared to the outer exchange couplings; thus, the low-energy global singlet state remains almost fully $|S_{12}S_{34}\rangle$, with $|S_{ij}\rangle = (1/\sqrt{2})(|\uparrow_i\downarrow_j\rangle - |\downarrow_i\uparrow_j\rangle)$. Figure 2(b) shows the result of a corresponding numerical simulation, which helps to interpret the data. The detunings for the anticrossing between the low-energy singlet and the $T_1^+$ are either vertical, where $T_1^+ \approx |T_{12}^+ S_{34}\rangle$ with $|T_{ij}^+\rangle = |\uparrow_i\uparrow_j\rangle$, or horizontal, where $T_1^+ \approx |S_{12}T_{34}^+\rangle$, over a large range of detunings. This demonstrates the independent control of $J_{12}$ and $J_{34}$ with the detunings $\varepsilon_{12}$ and $\varepsilon_{34}$, respectively. For higher $\varepsilon_{12}$ ($\varepsilon_{34}$), the horizontal (vertical) $T_0^+$ line bends away toward lower $\varepsilon_{34}$ ($\varepsilon_{12}$), which is a manifestation of the capacitive coupling between the left and right pair of dots: The singlet energy on one pair is lowered when the charge occupation for a singlet on the other pair becomes more (02)-like [34]. The capacitive coupling is modeled by adding $-DJ_{12}J_{34}$ to the diagonal matrix element for $|S_{12}S_{34}\rangle = |0_S\rangle$ in Eq. (3), with $D = 0.015$ $\mu eV^{-1}$ a prefactor for the interaction strength of the singlet dipoles [34,37]. At the left and right detunings for which the anticrossings with $T_0^+$ and $T_1^+$ are closest together, the condition $J_{12} = J_{34} = E_Z$ is reached, with $E_Z$ the Zeeman splitting set by the magnetic field.

The coupling between the singlet state and the quintuplet states is of second order in the hyperfine gradients; hence, the mixing between them is less efficient. The corresponding lines, such as the blue line in Fig. 2(b), are most visible when the quintuplet state energy is closest to a triplet state energy (see the white arrow), since the triplet states mediate the second-order coupling (see Appendix D).

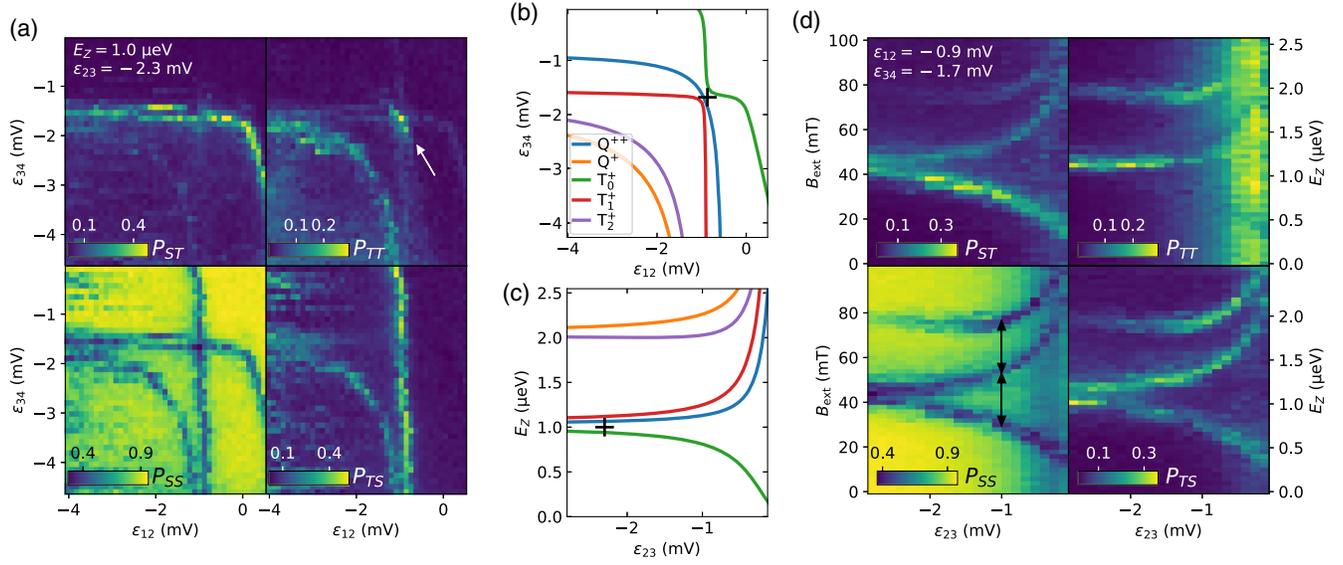

FIG. 2. Energy spectroscopy. Left- and right-pair correlated singlet-triplet probabilities from independent single-shot Pauli spin blockade readout as a function of (a) left- and right-pair detunings and (d) magnetic field and middle-pair detuning. A decrease in singlet-singlet probability and an increase in one of the other probabilities correspond to an anticrossing between the low-energy global singlet state and a polarized state. In (d), the right axis shows the calculated Zeeman splitting taking a Landé $g$ factor of $-0.44$. (b) Numerical simulation of the left- and right-pair detuning for which the low-energy global singlet state is degenerate with a polarized state. The parameters for the numerical simulation are obtained from separate spin funnel measurements for the left and right exchange coupling (see Appendix C 5) and from the Fourier transform in Fig. 3(d). (c) Numerical simulation of the middle-pair detuning and Zeeman splitting for which the low-energy global singlet state is degenerate with a polarized state. The diagram shows only the energies for the polarized states for which the magnetic field lowers the energy. The energy of the low-energy global singlet state is set to zero as a reference. Points in (b) and (c) with the same detunings and magnetic field are indicated with a "+". The legend for (c) is the same as for (b).

Figure 2(d) shows the measured energy-level diagram for which the middle-pair detuning and magnetic field are varied. The left- and right-pair detunings are fixed, and such that $J_{12} = J_{34} = E_Z$, as identified in Fig. 2(a). Figure 2(c) shows the corresponding numerical simulation. As $\varepsilon_{23}$ is increased, the energy splitting between $T_0^+$ and $T_1^+$ increases due to increased middle-pair exchange. We experimentally reach the condition $J_{12} = J_{23} = J_{34} = E_Z$ at the middle-pair detuning for which the energy-level spacing between $T_2^+$ and $T_1^+$ is equal to that for $T_1^+$ and $T_0^+$, which is indicated by the two equal-length double arrows in the bottom-left panel.

## V. GLOBAL COHERENT OSCILLATIONS

Before we further characterize the homogeneously coupled spin chain, we first demonstrate the coherent nature of the coupled four-spin system. Figure 3 shows global coherent oscillations, during which the full four-spin system evolves, along with Fourier transforms of those oscillations. Because of the symmetries of the Heisenberg Hamiltonian, time evolution occurs within the subspaces of fixed total spin and magnetization. Since we initialize in either only local singlets or at most one local triplet, the subspaces here consist of global singlet states or triplet states, respectively. The insets show numerical simulations based on time evolution under a single-band Fermi-Hubbard model [13] without decoherence effects (see Appendix E). We note that the condition of homogeneous exchange couplings can be extracted from the coherent oscillations as well.

To observe global coherent oscillations, the spin chain is again operated as depicted in Fig. 1(b). A magnetic field of 200 mT is applied here and in the subsequent measurements, to suppress leakage to states with different magnetization during the manipulation stage. For the data shown in Figs. 3(a)–3(d), a triplet state is initialized on the left pair and a singlet on the right pair. The detunings are rapidly pulsed such that the exchange couplings diabatically change and the system evolves coherently under the Heisenberg Hamiltonian during the manipulation stage. This evolution results in oscillations in the singlet-triplet probabilities of the left- and right-pair readout.

Figure 3(a) shows a chevron pattern from coherent oscillations between the global triplet states for varying differences between the left- and right-pair detuning and fixed middle-pair detuning. The Fourier transform of these oscillations is shown in Fig. 3(b). In line with the discussion in Sec. II, for fixed $J_{23}$ the energy difference between the two lowest-energy triplet states, and thus the oscillation frequency, is minimized if $J_{12} = J_{34}$. We point out that all three exchanges are activated in this measurement, so all four spins coherently evolve together.

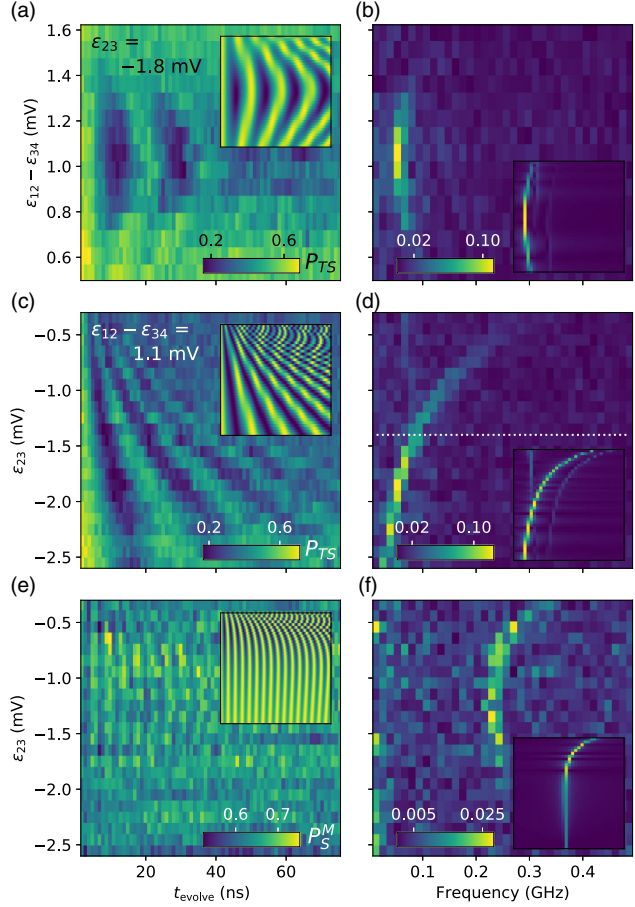

FIG. 3. Global coherent exchange oscillations. Coherent oscillations (a),(c) within the triplet subspaces and (e) within the singlet subspace. In (a), the difference in left- and right-pair detuning is varied, and in (c),(e), the middle-pair detuning is varied (with $J_{12} = J_{34}$). (b) Fourier transform of the data in (a). The frequency minimum corresponds to $J_{12} = J_{34}$. (d) Fourier transform of the data in (c). At $J_{12} = J_{23} = J_{34} = J_{\text{hom}}$, indicated by the white dotted line, the most visible frequency is equal to $J_{\text{hom}}/\sqrt{2}$. (f) Fourier transform of the data in (e). The frequency minimum corresponds to $J_{12} = J_{23} = J_{34} = J_{\text{hom}}$ and is equal to $\sqrt{3}J_{\text{hom}}$. Insets in all panels show numerical simulations of the experiment.

Figure 3(c) shows global coherent oscillations in the triplet subspace for varying middle-pair detuning and with the left- and right-pair detuning fixed such that $J_{12} = J_{34}$, as obtained from Figs. 3(a) and 3(b). In Fig. 3(d), the Fourier transform of these oscillations is shown. The middle-pair detuning for which $J_{12} = J_{23} = J_{34}$, indicated by the white dotted line, can be identified from this Fourier transform as the point where the faint vertical line meets the other more visible line. Here, the triplets with identical magnetization are equidistant in energy ($J_{\text{hom}}/\sqrt{2}$), thus reaching the condition of a spin chain with homogeneous exchange couplings, as described in Sec. II. In both the experimental and numerical Fourier transform data shown in Figs. 3(b) and 3(d), one frequency component is typically much more visible than the others. This is caused by the fact that the initial state overlaps mostly with just two of the three eigenstates; hence, the energy difference between these two eigenstates dominates the time evolution.

For the measurements in Fig. 3(e), a product of singlets is initialized, and the middle pair is read out. As described in Sec. II, the energy splitting in the singlet subspace, given $J_{12} = J_{34}$, is minimized when $J_{23} = J_{12} = J_{34}$. Figure 3(f) shows the Fourier transform of the oscillations in the singlet subspace. The $\varepsilon_{23}$ value for the frequency minimum cannot be precisely identified due to the limited frequency resolution, but an approximate identification is in agreement with the value of $\varepsilon_{23}$ corresponding to homogeneous exchange couplings from Fig. 3(d). Also, the ratio of the observed oscillation frequencies in the triplet and the singlet subspace at this value of $\varepsilon_{23}$ is consistent with theory.

The observation of global coherent oscillations demonstrates the coherent nature of the four-spin system. The coherence is limited by hyperfine and charge noise, of which the first can be strongly reduced by working with (isotopically purified) silicon or germanium as host materials [38]. The latter can be largely mitigated when simulating spin models at half filling (one electron per site) by operating at a so-called sweet spot [39,40]. Magnetic field gradients, such as due to hyperfine fields, can also induce leakage out of the fixed total spin and magnetization subspace. For the evolution, this can result in damping of the oscillations toward an offset that corresponds to the leakage state(s) [41], but this effect is strongly suppressed when exchange couplings dominate the hyperfine fields. The optimal visibility of the oscillations is, in general, lower than one, because the eigenstates partially overlap with the readout basis. The measured visibility is further lowered by relaxation during readout (which can be accounted for; see Appendix C 6), leakage, and partial adiabaticity of state preparation and transition to the readout configuration, which are further discussed in the next section.

## VI. PROBING THE LOW-ENERGY SINGLET

We finally turn to the preparation and characterization of the Heisenberg spin chain with homogeneous exchange couplings. The ground state of the Heisenberg spin chain, in the absence of an external magnetic field, is the low-energy singlet eigenstate. For homogeneous exchange couplings, $J_{ij} = J_{\text{hom}}$, the low-energy singlet eigenstate $S_0$, written in the singlet-triplet basis for the left and right pair, is

$$|S_0\rangle \propto (2\sqrt{3}+3)|S_{12}S_{34}\rangle + |T_{12}^0 T_{34}^0\rangle - |T_{12}^+ T_{34}^-\rangle - |T_{12}^- T_{34}^+\rangle, \tag{7}$$

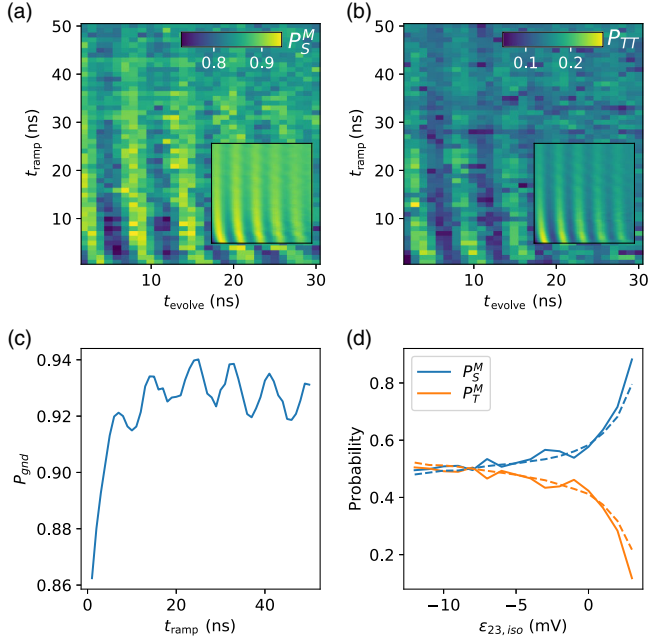

FIG. 4. Adiabaticity of state preparation and readout. (a) Singlet probability for middle-pair readout as a function of ramp-in time and evolution time. (b) Triplet-triplet probability for left- and right-pair readout as a function of ramp-in time and evolution time. Insets in (a) and (b) show numerical simulations of the experiment. (c) Numerical simulation of the overlap, $P_{\mathrm{gnd}} = |\langle \psi | S_0 \rangle|^2$, between the state at the manipulation stage, $|\psi\rangle$, and the low-energy singlet eigenstate $|S_0\rangle$, as a function of ramp-in time. The overlap is an indication for the success of the state preparation. (d) Middle-pair singlet and triplet probabilities from experiment (solid lines) and numerical simulations (dashed lines) as a function of middle-pair detuning for the isolation stage after the manipulation. The experimentally obtained probabilities are corrected for relaxation during the readout time (see Appendix C 6).

with normalization factor $1/2\sqrt{3(\sqrt{3}+2)}$. Upon measurement in the two-spin singlet-triplet basis for the left and right pair, we thus have a $\frac{1}{4}(2+\sqrt{3}) \approx 0.93$ singlet-singlet probability and approximately 0.07 triplet-triplet probability. Alternatively, the same global singlet state written in a basis given by the middle and outer pair is

$$|S_0\rangle = \frac{1}{\sqrt{2}} S_{14} S_{23} - \frac{1}{\sqrt{6}} T_{14}^0 T_{23}^0 + \frac{1}{\sqrt{6}} T_{14}^+ T_{23}^- + \frac{1}{\sqrt{6}} T_{14}^- T_{23}^+,$$

(8)

which indicates a 50:50 probability to measure a singlet or a triplet on the middle pair. When quasistatic hyperfine and charge noise is included (see Appendix E), then numerical simulations result in probabilities of $P_S^M = 0.50$, $P_T^M = 0.50$, $P_{SS} = 0.91$, $P_{ST} = 0.01$, $P_{TS} = 0.01$, and $P_{TT} = 0.07$, which indicates that the noise in the device does not form a direct bottleneck for the quantitative characterization of the spin-chain ground state.

Quantitative two-spin singlet-triplet characterization of the low-energy singlet eigenstate $S_0$ is facilitated by state preparation that is adiabatic with respect to the exchange couplings. Starting from singlets on the left and right pair of dots, the detunings are slowly varied using voltage ramps to increase the middle-pair exchange while reducing the left- and right-pair exchange. Ideally, the singlet product state evolves to the instantaneous low-energy singlet eigenstate of the Hamiltonian at the manipulation stage. Conversely, projection in the singlet-triplet basis for readout requires a diabatic transition from the manipulation to the readout stage.

Figure 4(a) shows the singlet probability for the middle-pair readout, and Fig. 4(b) shows the triplet-triplet probability for the left- and right-pair readout as a function of ramp-in time and evolution time. As expected, we observe oscillations with a decreasing visibility as the ramp-in time increases, as a result of the more adiabatic character of the state preparation. We note that, for the measurements shown in Figs. 4(a)–4(c) during the manipulation stage, the middle exchange $J_{23} \approx 175$ MHz is made larger than the outer exchanges, $J_{12} = J_{34} \approx 85$ MHz, in order to increase the visibility of the oscillations, so we can best evaluate the adiabaticity of the state preparation. For shorter ramp-in time, the oscillations bend toward longer evolution time, which is due to the evolution during the ramp-in; thus, at the start of the evolution stage, the state has already evolved further for longer ramp-in time. Figure 4(c) shows the numerically simulated overlap with the low-energy singlet eigenstate at the manipulation point as a function of ramp-in time, indicating >92% overlap for 12–36 ns ramp-in time. For longer ramp-in time, leakage to quintuplet and triplet states starts to dominate.

We next study how to maximize the degree of diabaticity for the readout pulses with respect to the exchange couplings, in order to acquire the singlet-triplet probabilities for the state at the end of the manipulation stage. To increase the diabaticity given the finite rise time of the arbitrary waveform generator, an isolation stage is added between the manipulation stage and the readout stage for the case where we aim to read the middle pair. In the isolation stage, the voltages are pulsed deep into the (1201) charge region, such that the voltage step is steeper, which makes the pulse more diabatic.

Figure 4(d) shows the singlet and triplet probability for the middle-pair readout as a function of $\varepsilon_{23,\mathrm{iso}}$, the middle-pair detuning (relative to the readout position) for the isolation stage. The state is prepared with $t_{\mathrm{ramp}} = 25$ ns and with homogeneous exchange couplings at the manipulation stage. As discussed previously, for this condition the low-energy singlet eigenstate ideally has equal two-spin singlet and triplet probability on the middle pair. We see in Fig. 4(d) that the measured two-spin singlet and triplet probabilities gradually approach each other as $\varepsilon_{23,\text{iso}}$ is made more negative, pulsing deeper in the (1201) charge region, indicating that the isolation becomes more diabatic in exchange couplings. Pushing $\varepsilon_{23,\text{iso}}$ even further, the singlet-triplet probabilities for the middle pair pass slightly beyond 50/50%, which we can trace back to an artifact from the digital filter in the arbitrary waveform generator. Specifically, we measure the detailed rising flank of the arbitrary waveform generator, which shows an undershoot just before the rising flank and ringing after the rising flank, and numerically simulate its effect on the measured singlet-triplet probabilities (see Appendix E).

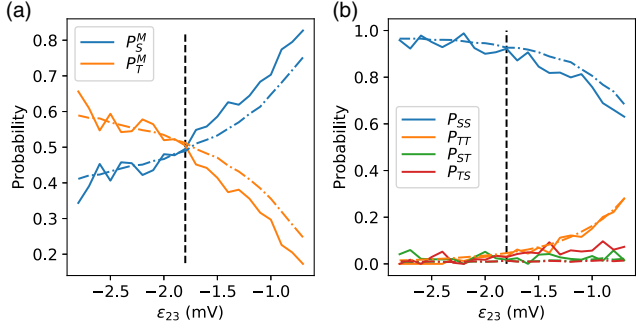

FIG. 5. Probing the low-energy singlet eigenstate of the homogeneous spin chain. (a) Singlet and triplet probability for middle-pair readout and (b) singlet-triplet correlated probabilities from left- and right-pair readout as a function of middle-pair detuning and with $J_{12} = J_{34}$. Experimental data are shown as solid lines and the results from numerical simulations as dashed-dotted lines. Vertical dashed lines indicate the middle-pair detuning for which the exchange couplings are calibrated to be homogeneous. The experimentally obtained probabilities are corrected for relaxation during the readout time (see Appendix C 6).

Based on these findings, we set in what follows the ramp-in time to 26 ns when aiming to adiabatically prepare the low-energy singlet eigenstate. Numerical simulations similar to Fig. 4(c) show that an overlap of 95.5% with the low-energy singlet state is expected for the homogeneous spin chain. For readout of the middle pair, the middle-pair isolation detuning is set to $\varepsilon_{23,\text{iso}} = 10$ mV.

Figure 5 shows the singlet-triplet probabilities for the spin chain as a function of middle-pair detuning (with fixed $J_{12} = J_{34} \approx 85$ MHz) during the manipulation stage. The exchange couplings are calibrated to be homogeneous at $\varepsilon_{23} = -1.8$ mV based on measurements of coherent oscillations similar to Fig. 3. The exchange coupling favors two-spin singlet formation on every pair, but, due to the monogamy of entanglement, there cannot simultaneously exist such singlets on overlapping pairs. Away from homogeneous exchange couplings, we see in Fig. 5(a) that, for increasing $\varepsilon_{23}$ and, thus, increasing $J_{23}$, the singlet probability on the middle pair increases, as expected. Conversely, in Fig. 5(b), we observe that as $\varepsilon_{23}$ decreases ($J_{23}$ decreases), $P_{SS}$ ($P_{TT}$) increases (decreases), because the singlets on the left and right pair become energetically increasingly favorable compared to the middle-pair singlet. For higher $\varepsilon_{23}$, also $P_{TS}$ and $P_{ST}$ increase, which is caused by leakage out of the singlet subspace, due to decreasing energy splitting between the global singlet states and other $m_S = 0$ states. The experimentally measured numbers are in good agreement with the predicted probabilities based on Eqs. (7) and (8) and numerical simulations including noise. Both the qualitative trend and quantitative comparison between experiment and simulation indicate high-fidelity preparation of the low-energy singlet eigenstate for the homogeneously coupled spin chain and a high-fidelity characterization method based on singlet-triplet readout and correlations therein.

## VII. CONCLUSION AND OUTLOOK

In summary, we have implemented a quantum simulation of a quantum coherent antiferromagnetic Heisenberg spin chain. For this purpose, we have developed energy spectroscopy methods to identify the condition of homogeneous exchange couplings. Furthermore, we have devised methods to prepare the global ground state of the homogeneous Heisenberg spin-chain Hamiltonian and to probe this state via local measurements in the singlet-triplet basis and correlations of such measurements. We find both qualitative and quantitative agreement between experiment and numerical simulation. Finally, coherent oscillations of the full four-spin system indicate the coherent nature of the system, despite the presence of hyperfine noise in the GaAs host material.

Future quantum magnetism simulation experiments with quantum dots may leverage the recent developments of (isotopically purified) silicon and germanium as host materials, due to the lower concentration of nuclear spins, which further enhances coherence and facilitates high-resolution spectroscopy. The demonstrated control of exchange couplings, as facilitated by the independent control with virtual gates, is a powerful technique for quantum simulations in larger systems. The techniques demonstrated here also pave the way for quantum magnetism simulations in other lattice configurations, such as square and triangular ladders for which simulations of, respectively, resonating valence bonds and frustrated magnetism are now within the capabilities of gate-defined quantum dots.

The data reported in this paper and scripts to generate the figures are uploaded to Zenodo [42].

## ACKNOWLEDGMENTS

The authors thank C. E. Bos, J. P. Dehollain, T. Fujita, S. P. M. van Gemert, T. Hensgens, V. P. Michal, and the other members of the Vandersypen group for stimulating discussions and acknowledge software development by

S. L. de Snoo and technical support by O. Benninghof and R. van Leeuwen. This work was supported by the Dutch Research Council (NWO Vici), European Research Council (ERC Advanced Grant), and the Swiss National Science Foundation.

## APPENDIX A: LIMITATION FOR DISTINGUISHING STATES WITH TWO-SPIN SINGLET-TRIPLET MEASUREMENTS

The two-spin singlet and triplet projection operators are, respectively,

$$P_{S_{ij}} = \frac{1}{4} - \vec{S}_i \cdot \vec{S}_j, \tag{A1}$$

$$P_{T_{ij}} = 1 - P_{S_{ij}}. \tag{A2}$$

The global spin-raising operator $\hat{S}^+ = \sum_i \hat{S}_i^+$, with $\hat{S}_i^+ = \hat{S}_i^x + i\hat{S}_i^y$, commutes with the singlet and triplet projection operators.

In addition, the spin-raising operator commutes with the Heisenberg Hamiltonian

$$[H_{\text{Heis}}, \hat{S}^+] = 0; \tag{A3}$$

thus, the spin-raising and -lowering operators map between states in the same eigenspace of the Hamiltonian.

From the commutativity, it follows that two-spin singlet-triplet measurements cannot distinguish states in the same eigenspace for the Heisenberg Hamiltonian, and those states can be mapped onto one another with spin-raising or -lowering operators.

## APPENDIX B: GLOBAL TRIPLET ENERGIES

Figure 6 shows the triplet energies as a function of $J_{12} - J_{34}$ and as a function of $J_{23}$ with $J_{12} = J_{34}$.

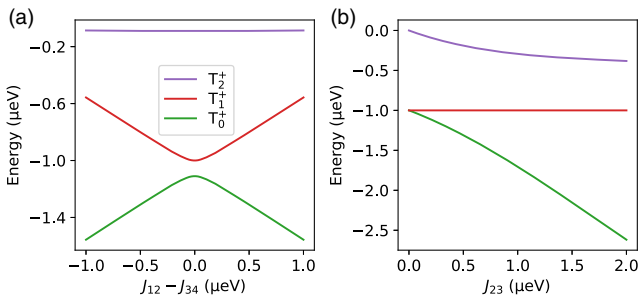

FIG. 6. Triplet energies. (a) Energies of global triplet states as a function of $J_{12} - J_{34}$, with $J_{23} = 0.2$ $\mu$eV and $J_{12} + J_{34} = 2$ $\mu$eV, revealing a minimum in energy difference at $J_{12} = J_{34}$ for $T_0^+$ and $T_1^+$. (b) Triplet energies as a function of $J_{23}$ with $J_{12} = J_{34} = 1$ $\mu$eV, showing that at $J_{12} = J_{23} = J_{34}$ the energy spacings are equal. The legend for (b) is the same as for (a). An external magnetic field results in an overall offset and, thus, does not change the triplet energy differences.

## APPENDIX C: EXPERIMENTAL METHODS

### 1. Device

The quadruple quantum dot and sensing dot are formed in a device designed for eight dots and two sensors. A scanning electron micrograph image of the active region of a device similar to the one used in the experiment is shown in Fig. 1(a). The device is mounted in a dilution refrigerator, which results in an electron reservoir temperature of about 100 mK (roughly 10 $\mu$eV). By applying voltages on the electrodes on the surface, we shape the potential landscape in a two-dimensional electron gas 90 nm below the surface, formed in a silicon-doped GaAs/AlGaAs heterostructure. The plunger gates, labeled $P_i$ [red in Fig. 1(a)] for the spin-chain dots and SDP for the sensing dot, control the electrochemical potentials, and the barrier gates [green in Fig. 1(a)] control the tunnel couplings between dots or between a dot and a reservoir. When an external magnetic field is applied, it is oriented in the plane of the 2D electron gas.

Table II shows an overview of the pulse sequence, with the durations and descriptions of the pulse stages.

### 2. Virtual gates

For the independent control of site-specific offsets [43], the cross talk is compensated with the matrix transformation

$$\begin{pmatrix} \varepsilon_1 \\ \varepsilon_2 \\ \varepsilon_3 \\ \varepsilon_4 \end{pmatrix} = \begin{pmatrix} 1 & 0.539 & 0.203 & 0.145 \\ 0.542 & 1 & 0.538 & 0.223 \\ 0.181 & 0.5 & 1 & 0.507 \\ 0.100 & 0.242 & 0.522 & 1 \end{pmatrix} \begin{pmatrix} P_1 \\ P_2 \\ P_3 \\ P_4 \end{pmatrix}. \tag{C1}$$

The lever arms for the single-particle energy offsets are [76, 81, 87, 84] $\mu$eV mV$^{-1}$, which are obtained with photon-assisted tunneling [31,44].

### 3. Charge-stability diagrams with sequence details

Figure 7 shows charge-stability diagrams for the nearest-neighbor pairs in the quadruple dot. Typical pulse voltage positions and the order of the pulse sequence are indicated. The compensation and the parking for the readout of the left and right pair are not displayed to preserve clarity.

### 4. Postselection for state preparation

Figure 8 shows single-shot results of readout directly after initialization. The data for the readout after the evolution stage are postselected by thresholding the signal from the readout after initialization. The signal is also used for background subtraction to suppress the effect of low-frequency charge noise on the signal for the final readout. Errors in initialization for both the singlet-singlet product state and the triplet-singlet product state can occur because of the finite dot-reservoir tunnel rate, due to which the target charge state (0202) and (1102), respectively, is not occupied at the end of the initialization stage. These errors result in the counts in the top half regions in Figs. 8(a) and 8(b). Such errors can be reduced by increasing the duration of the initialization stage or increasing the dot-reservoir tunnel rate. Additionally, errors in the triplet-singlet product initialization can be caused by thermal excitations, due to which a singlet-singlet product state in the (1102) charge occupation can be occupied, which results in the bottom-left peak in Fig. 8(b). This error can be reduced by increasing the magnetic field strength.

TABLE II. Details on the stages for the pulse sequence with the durations and descriptions as used for the spin-chain operation. The total duration of the pulse sequence is below 191 $\mu s$, including the compensation stage.

| Stage | Duration | Description |
|---|---|---|
| Initialization ($L$) | 100 $\mu s$ | Exchange electrons with the reservoir, either in the (0202) charge region for a singlet-singlet state or in (1102) for a triplet-singlet state |
| Readout initial ($R$) | 20 $\mu s$ | Readout of the left and the right pair (each 10 $\mu s$) for postselection and for background signal subtraction for the final readout to correct for low-frequency charge noise |
| Separate ($S$) | 2 ns | Separate electrons into (1111) charge region |
| Ramp detunings | 0–50 ns | Ramp the voltages to the operation point ($O$) |
| Evolve ($O$) | 2–100 ns | Let the spin state evolve under the exchange couplings |
| Isolate ($I/S$) | 2 ns | Isolate the pair(s) for readout |
| Readout final ($R$) | 20 $\mu s$ | Read out the middle pair or the left and right pair (each 10 $\mu s$) |
| Compensate | <50 $\mu s$ | Discharge the capacitors in the bias tee's |

### 5. Separate exchange coupling measurements

Figure 9 shows separate exchange coupling measurements for each of the neighboring pairs. For the spin funnel measurements for the left and right pair, the middle exchange coupling is set to be small. The spin funnels are fitted with the exchange model in Eq. (6). For the middle pair, the Fourier transform of coherent oscillations in the triplet-subspace is used. The model for the Fourier transform follows from Eq. (4) and is $\frac{1}{2}(J_{23} - J_{hom} + \sqrt{J_{23}^2 + J_{hom}^2})$, with $J_{hom} = J_{12} = J_{34} = 125$ MHZ, the homogeneous exchange coupling as obtained from Fig. 3. From these three separate fits, the tunnel couplings are extracted as 8.5, 7.5, and 11.9 $\mu$eV for the left, middle, and right pair, respectively. These tunnel coupling values are used for the numerical simulations of the experiment.

### 6. Readout with relaxation and histogram models

Single-shot readout characteristics for each of the nearest-neighbor pairs are shown in Fig. 10. The data in Figs. 10(b) and 10(d) are from the data for Fig. 5(a), and the data from Fig. 10(e) is from part of the data for Fig. 2(a).

The effect on the probabilities of relaxation during the readout is taken into account by modeling the single-shot data with a histogram. For readout on a single pair, the model for the histogram is described in Refs. [33,45]. For

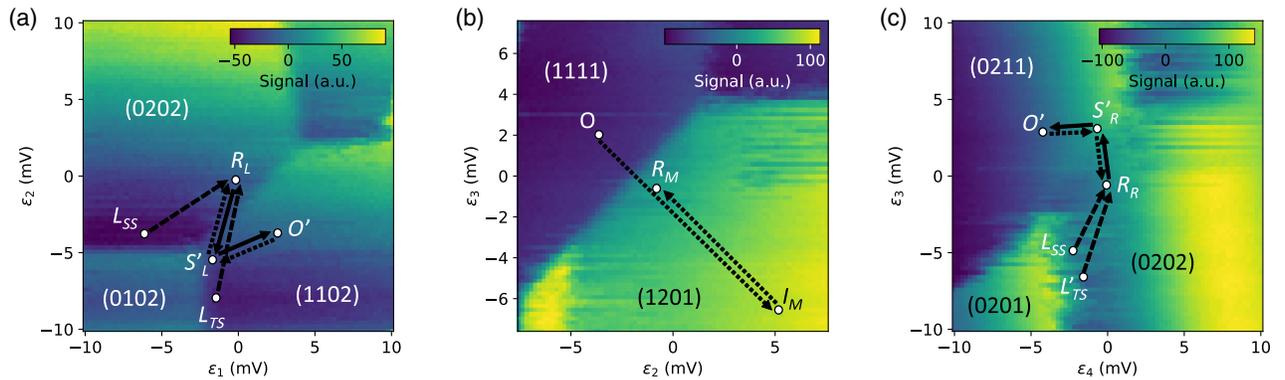

FIG. 7. Charge-stability diagrams for nearest-neighbor pairs, with (a) left, (b) middle, and (c) right. Pulse positions are indicated with small white circles and are labeled with a letter corresponding to the pulse stage in Table II. Primes are added to characters for which the indicated voltage position is a projection, because the true position is in a different charge occupation, which is outside the respective two-dimensional plane in the four-dimensional charge-stability space. Solid arrows represent the part of the pulse sequence for which only one variant is used, though the operation voltage position $O$ is varied throughout the experiments. Dashed arrows correspond to sequences for initialization of either a singlet-singlet $L_{SS}$ or triplet-singlet $L_{TS}$. Dotted arrows correspond to sequences for readout (and isolation) of either the middle pair, $R_M$ ($I_M$), or readout (and isolation) of the left pair, $R_L$ ($S'_L$), and the right pair, $R_R$ ($S'_R$).

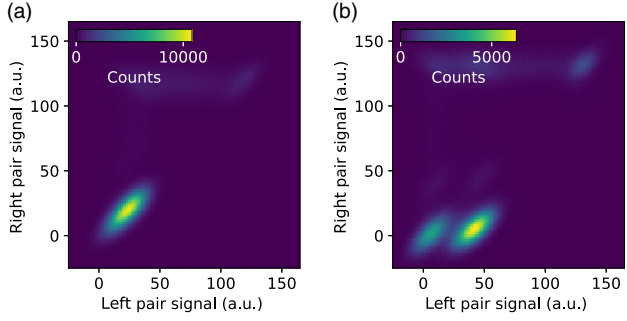

FIG. 8. Histograms of correlated left- and right-pair readout directly after initialization of (a) the singlet-singlet product state and (b) the triplet-singlet product state. For both the left- and the right-pair readout, the integration time is 8 $\mu$s, and the bin width is [1.9, 1.9] a.u.

the singlet-triplet correlation measurements, the two-dimensional single-shot histograms are modeled with

$$N(\mathbf{x}) = N_{\text{tot}} \sum_{i,j \in [S,T]} P_{ij} n_{ij}(\mathbf{x}) w_1 w_2, \quad \text{(C2)}$$

with $N_{\text{tot}}$ the number of single-shot repetitions, $P_{ij}$ the average probability for outcome $ij$, $n_{ij}$ the probability density distribution, and $w_i$ the bin widths.

The readout of the left and right pair of spins is performed with the same sensing dot; thus, noise on the sensor signal can induce correlations between the two readouts. Low-frequency noise, which causes signal differences between repetitions, is taken into account by the subtraction of the sensor signal from the readout directly after the initialization. Correlations in the sensor signals due to higher-frequency noise remain, which results in the diagonally elongated signal peaks in the two-dimensional readout histogram as shown in Fig. 10(e). The effect of the correlations is incorporated in the model for the two-dimensional Gaussian histogram by modeling the Gaussian peaks with rotated ellipses.

The two-dimensional Gaussian is

$$g_{2D}(\mathbf{x}, \boldsymbol{\mu}) = \frac{1}{2\pi\sigma_1\sigma_2} e^{-[a(x_1-\mu_1)^2 + 2b(x_1-\mu_1)(x_2-\mu_2) + c(x_2-\mu_2)^2]}, \quad \text{(C3)}$$

with $\boldsymbol{\mu}$ the mean coordinates and where the parameters for the shapes of the Gaussian peaks are

$$a = \frac{\cos^2\theta}{2\sigma_1^2} + \frac{\sin^2\theta}{2\sigma_2^2}, \qquad b = -\frac{\sin 2\theta}{4\sigma_1^2} + \frac{\sin 2\theta}{4\sigma_2^2},$$
$$c = \frac{\sin^2\theta}{2\sigma_1^2} + \frac{\cos^2\theta}{2\sigma_2^2}, \quad \text{(C4)}$$

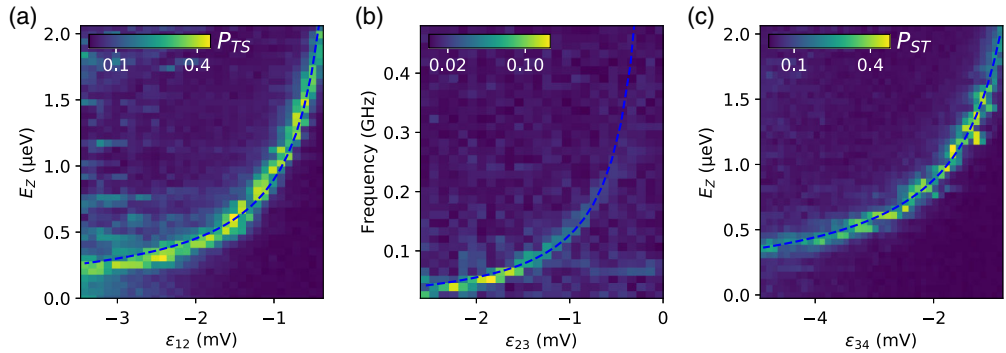

FIG. 9. Separate exchange coupling measurements. The dashed blue lines are fits to the data. (a) Spin funnel on left pair, (b) Fourier transform of triplet-subspace coherent oscillations shown in Fig. 3(c), (c) spin funnel on right pair.

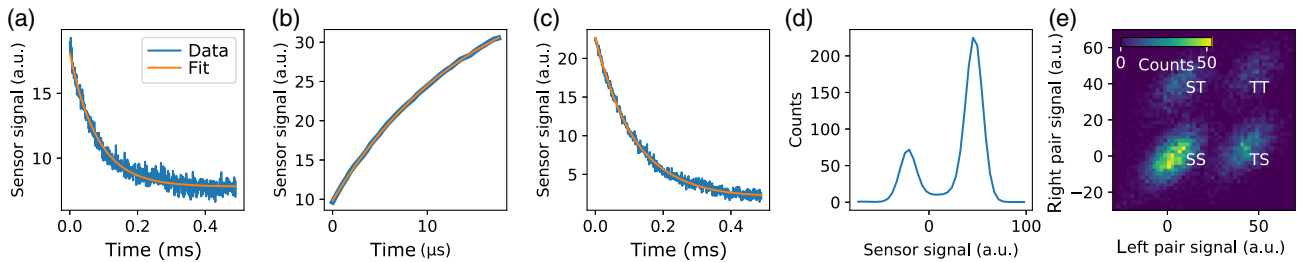

FIG. 10. Readout characteristics (a)–(c) show relaxation curves for the Pauli spin blockade readout of the left, middle, and right pair, respectively. The corresponding $T_1$ values are 80.9, 25.0, and 114.8 $\mu$s. (d) Histogram of middle-pair readout with 8 $\mu$s integration time and bin width 4.375 a.u., and the counts are in thousands. (e) Histogram of correlated left- and right-pair readout. For both readouts, the integration time is 8 $\mu$s, and the bin width is [2.0, 2.0] a.u.

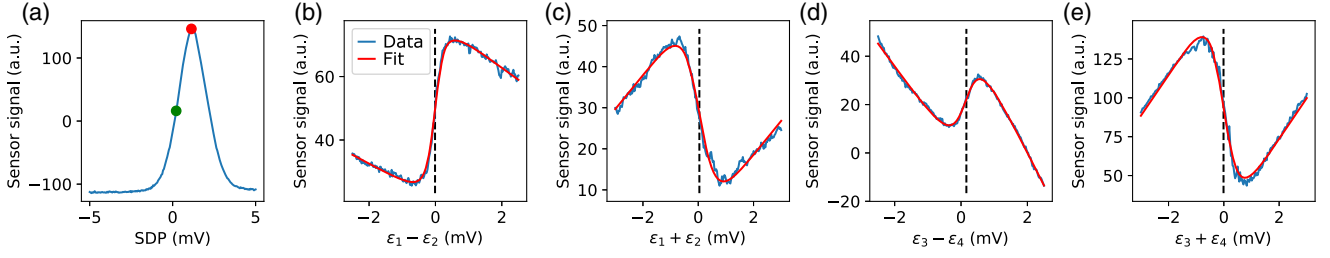

FIG. 11. Automatic calibration of pulse voltage offsets. (a) Sensing dot calibration, with the red marker corresponding to the top of the Coulomb peak, and the green marker is the half-height on the rising flank. (b) interdot transition between (0202) and (1102), (c) dot-reservoir transition between (0202) and (1202), (d) interdot transition between (0202) and (0211), (e) dot-reservoir transition between (0202) and (0212). The traces are not centered at the same voltages, but voltage offsets are added for each scan based on the previous calibration.

with $\theta$ the rotation angle of the ellipsoidal shape of the Gaussian peaks, and $\sigma_1$ and $\sigma_2$ describe the ellipse width and length. In the experiment, the first and second readout signals are integrated for equal durations, which sets $\theta = \pi/4$, which means that the one-dimensional histograms for the left- and right-pair readout have the same Gaussian widths. In the extreme when there are no noise correlations, then $\sigma_1 = \sigma_2$, and, in the other extreme where the noise would be fully correlated, the histogram effectively is a one-dimensional Gaussian.

The probability distributions for the correlated singlet-triplet outcomes are

$$n_{SS}(\mathbf{x}) = g_{2D}(\mathbf{x}, \boldsymbol{\mu}_{SS}), \tag{C5}$$

$$n_{ST}(\mathbf{x}) = e^{-\alpha_R} g_{2D}(\mathbf{x}, \boldsymbol{\mu}_{ST}) + \alpha_R \int_0^1 dz e^{-z\alpha_R} g_{2D}(\mathbf{x}, z(\boldsymbol{\mu}_{ST} - \boldsymbol{\mu}_{SS}) + \boldsymbol{\mu}_{SS}), \tag{C6}$$

$$n_{TS}(\mathbf{x}) = e^{-\alpha_L} g_{2D}(\mathbf{x}, \boldsymbol{\mu}_{TS}) + \alpha_L \int_0^1 dz e^{-z\alpha_L} g_{2D}(\mathbf{x}, z(\boldsymbol{\mu}_{TS} - \boldsymbol{\mu}_{SS}) + \boldsymbol{\mu}_{SS}), \tag{C7}$$

$$\begin{aligned} n_{TT}(\mathbf{x}) = {}& e^{-(\alpha_L + \alpha_R)} g_{2D}(\mathbf{x}, \boldsymbol{\mu}_{TT}) + e^{-\alpha_R} \alpha_L \int_0^1 dz e^{-z\alpha_L} g_{2D}(\mathbf{x}, z(\boldsymbol{\mu}_{TT} - \boldsymbol{\mu}_{ST}) + \boldsymbol{\mu}_{ST}) \\ & + e^{-\alpha_L} \alpha_R \int_0^1 dz e^{-z\alpha_R} g_{2D}(\mathbf{x}, z(\boldsymbol{\mu}_{TT} - \boldsymbol{\mu}_{TS}) + \boldsymbol{\mu}_{TS}) \\ & + \alpha_L \alpha_R \int_0^1 \int_0^1 dz dz' \left[ e^{-(z\alpha_L + z'\alpha_R)} g_{2D}\left(\mathbf{x}, (z(\mu_{T,L} - \mu_{S,L}) + \mu_{S,L}, z'(\mu_{T,R} - \mu_{S,R}) + \mu_{S,R})\right) \right], \end{aligned} \tag{C8}$$

with $\alpha_i = t_i/T_{1,i}$, which is the ratio between the signal integration time $t_i$ and the relaxation time $T_{1,i}$ for the left pair or the right pair. In $n_{TT}$, the first term corresponds to states which do not decay during both readouts, the second term to states which decay during the left-pair readout but not the right-pair readout, the third term to states which do not during the left-pair readout but do decay during the right-pair readout, and the last term to states which decay during both readouts.

For the fitting procedure, first the histogram of single-shot results for all pulse set points is fitted to obtain the Gaussian widths and peak positions, and then the histogram for each pulse set point is separately fitted to obtain the probability amplitudes.

## 7. Automatic pulse offset calibration

The spin-chain operation of the device is automatically recalibrated on a daily basis to correct for the effect of irregular charge jumps. The results of the calibration are incorporated as offsets for the pulse voltages. The static voltage on the sensing dot plunger is sometimes adjusted between measurements, but all other dc voltages are left untouched to avoid introducing instabilities of the device.

Figure 11 shows the result of the daily automatic recalibration, which consists of five traces. These traces are a sweep of the sensing dot plunger gate voltage and four one-dimensional cuts through the charge-stability space of the quadruple quantum dot. The sensing dot scan is performed to calibrate the gate voltage for the position on the flank of the sensing dot Coulomb peak. The four one-dimensional cuts are performed to locate the gate voltages corresponding to specific charge transitions. The gate voltages for these transitions form a reference for the voltages used in the pulsed control of the spin chain, such that electrostatic shifts of the device are corrected for.

The analysis procedures for the Coulomb peak and the dot-reservoir transition are based on Ref. [46], and the interdot transition fitting is from Ref. [44]. The automatic calibration routine takes in total approximately 30 s. Occasionally, large charge jumps require coarse manual tuning of the device, after which the automatic calibration routine is used for refinement:

## APPENDIX D: HYPERFINE GRADIENT COUPLINGS

The effect of the nuclear spins in the material environment on the electron spins in the quantum dots can effectively be described by hyperfine fields [32] as

$$H_{hf} = g\mu_B \sum_i \vec{h}_i \cdot \vec{S}_i, \tag{D1}$$

with $\vec{h}_i$ the local hyperfine field for dot $i$. The hyperfine term breaks the conservation of total spin and spin $z$ of the Heisenberg Hamiltonian and has been studied extensively experimentally and theoretically for two-spin systems [32,47]. For the energy spectroscopy of the four-spin system as shown in Fig. 2, the couplings between the singlet-subspace and the polarized states are given by the hyperfine matrix, which in the basis $(Q^{++}, Q^+, |2_{T^+}\rangle, |1_{T^+}\rangle, |0_{T^+}\rangle, |1_S\rangle, |0_S\rangle)$ is

$$
\begin{pmatrix}
\frac{B_{z,sum}}{2} & \frac{B_{x,sum}-iB_{y,sum}}{4} & \frac{dB_{-,14}+dB_{-,23}}{2} & \frac{-dB_{-,12}}{\sqrt{2}} & \frac{-dB_{-,34}}{\sqrt{2}} & 0 & 0 \\
\frac{B_{x,sum}+iB_{y,sum}}{4} & \frac{B_{z,sum}}{4} & \frac{-dB_{z,12}-dB_{z,34}}{2} & \frac{dB_{z,12}}{\sqrt{2}} & \frac{dB_{z,34}}{\sqrt{2}} & 0 & 0 \\
\frac{dB_{+,14}+dB_{+,23}}{2} & \frac{-dB_{z,12}-dB_{z,34}}{2} & \frac{B_{z,sum}}{4} & \frac{dB_{z,12}}{\sqrt{2}} & \frac{-dB_{z,12}}{\sqrt{2}} & \frac{dB_{-,14}+dB_{-,23}}{\sqrt{3}} & 0 \\
\frac{-dB_{+,12}}{\sqrt{2}} & \frac{dB_{z,12}}{\sqrt{2}} & \frac{dB_{z,12}}{\sqrt{2}} & \frac{B_{z,3}+B_{z,4}}{2} & 0 & \frac{dB_{-,12}}{\sqrt{6}} & \frac{-dB_{-,34}}{\sqrt{2}} \\
\frac{-dB_{+,34}}{\sqrt{2}} & \frac{dB_{z,34}}{\sqrt{2}} & \frac{-dB_{z,34}}{\sqrt{2}} & 0 & \frac{B_{z,1}+B_{z,2}}{2} & \frac{dB_{-,34}}{\sqrt{2}} & \frac{-dB_{-,12}}{\sqrt{6}} \\
0 & 0 & \frac{dB_{+,14}+dB_{+,23}}{\sqrt{3}} & \frac{dB_{+,12}}{\sqrt{6}} & \frac{dB_{+,34}}{\sqrt{2}} & 0 & 0 \\
0 & 0 & 0 & \frac{-dB_{+,34}}{\sqrt{2}} & \frac{-dB_{+,12}}{\sqrt{6}} & 0 & 0
\end{pmatrix}, \tag{D2}
$$

with $dB_{+,ij} = dB_{x,ij} + idB_{y,ij}$ and $dB_{-,ij} = dB_{x,ij} - idB_{y,ij}$, where $dB_{\alpha,ij} = (B_{\alpha,i} - B_{\alpha,j})/2$, and $B_{\alpha,sum} = \sum_i B_{\alpha,i}$.

From the hyperfine matrix, it follows that, for any state in the singlet subspace $|S_k\rangle$, the first-order coupling to a quintuplet is $\langle S_k|H_{hf}|Q^{++}\rangle = \langle S_k|H_{hf}|Q^+\rangle = 0$. The coupling between the singlet and the quintuplet(s), as visible in the energy spectroscopy in Fig. 2, can be explained by a second-order effect where the singlet state couples to a triplet state, which, in turn, couples to a quintuplet state. The second-order coupling becomes effective only when the singlet and quintuplet state energies are near the energy of a coupling-mediating triplet state. This is observed in the energy spectroscopy where the coupling of the singlet to the $Q^{++}$ state decreases as the singlet energy differs more from the $T_0^+$ and $T_1^+$ energies.

Note that the spin-orbit interaction can induce couplings between states with different charge occupation [48,49], which can contribute to the signal for the anticrossings in the energy spectroscopy.

## APPENDIX E: NUMERICAL SIMULATION

The time evolution simulations are based on the single-band Fermi-Hubbard model

$$
\begin{aligned}
H_{FH} = &-\sum_i \epsilon_i n_i + \sum_i \frac{U_i}{2} n_i(n_i - 1) \\
&+ \sum_{ij, i\neq j} V_{ij} n_i n_j - \sum_{\langle i,j \rangle} t_{ij}(c_i^\dagger c_j + \text{H.c.}),
\end{aligned}
$$

where $\epsilon_i$ is the negative single-particle energy offset, $n_i = c_i^\dagger c_i$ is the dot occupation, $c_i^{(\dagger)}$ is the annihilation (creation) operator, $U_i$ is the on-site Coulomb repulsion, $V_{ij}$ is the intersite Coulomb repulsion, and $t_{ij}$ is the interdot tunnel coupling. The parameter values for the simulations are $U_i = 3$ meV, $V_{i,i+1} = 0.5$ meV, $V_{i,i+2} = 0.1$ meV, $V_{i,i+3} = 20\ \mu$eV, and $[t_{12}, t_{23}, t_{34}] = [8.5, 7.5, 11.9]\ \mu$eV. The tunnel coupling values are experimentally obtained from fits to spin funnels for the left and right pair and from a fit of the Fourier transform in Fig. 3(d) (see Appendix C 5). Differences between the experimental and simulation values for the interaction parameters are accounted for in the simulations with offsets in the single-particle energies. The Hamiltonian matrix for the simulation is generated with QuTiP [50]. The charge configurations used for the simulations are (1111), (0211), (1201), (1102), and (0202). The spin subspace which is used for the simulation in Figs. 3(a)–3(d) is the four-spin $T^+$ subspace, while for Figs. 3(e) and 3(f) the global singlet subspace is used. For the numerical simulations in Figs. 4 and 5, the set of states for the simulation consists of the global singlet subspace, and all other $m_S = 0$ states (the four-spin $T^0$ subspace and $Q^0$).

For the simulation, the same sequence is followed as the experimental sequence for the spin chain, which is shown in Fig. 1(b) and detailed in Table II. Time evolution of the states is computed using an in-house density matrix solver package [51]. The simulated probabilities shown in the figures in the main text are the probabilities at the isolation stage after the evolution. In order to simulate the adiabaticity of the experimental sequence, the voltage pulse shape for the simulation is based on the experimental voltage pulse shape from the arbitrary waveform generator (AWG). The simulated shape is obtained from the Fourier transform of the ideal pulse shapes and the subsequent inverse Fourier transform with the experimentally measured frequency response of the AWG. The Fourier component amplitudes are corrected for the finite sampling rate $f_s = 1$ GHz of the AWG by dividing with $\mathrm{sinc}(f/f_s)$, where $f$ is the frequency of the Fourier component.

The effects of charge noise and quasistatic hyperfine noise are included for the simulations in the main text, except for Fig. 3. The quasistatic hyperfine noise is considered by repeating the simulation and for each repetition taking a sample from a Gaussian distribution with a root mean square of 3.2 mT, which is experimentally obtained from the amplitude decay of the global exchange oscillations. The power spectral density of the charge noise is modeled with $A/f^\alpha$, with $A = 0.26\ \mu\mathrm{eV}^2/\mathrm{Hz}$ and $\alpha = 0.79$, which are obtained from a charge-noise measurement. The charge noise is measured from 1 Hz to 5 kHz. For the simulations, a charge-noise frequency range from 0.1 Hz to 100 GHz is used, where noise on a timescale longer than a single shot of the experimental sequence is integrated and added as quasistatic noise.

---